# A molecular Ferroelectric thin film of imidazolium perchlorate on Silicon

Congqin Zheng, Xin Li, Yuhui Huang, Yongjun Wu[*], Zijian Hong[*]

[1]School of Materials Science and Engineering, Zhejiang University, Hangzhou 310027, China

[2]State Key Laboratory of Silicon and Advanced Semiconductor Materials, Zhejiang University, Hangzhou, Zhejiang 310027, China

*Email: yongjunwu@zju.edu.cn (YW); hongzijian100@zju.edu.cn (ZH)

**Abstract**

Molecular ferroelectric materials have attracted widespread attention due to their abundant chemical diversity, structural tunability, low synthesis temperature, and high flexibility. Meanwhile, the integration of molecular ferroelectric materials and Si is still challenging, while the fundamental understanding of the ferroelectric switching process is still lacking. Herein, we have successfully synthesized the imidazole perchlorate ($ImClO_4$) single crystals and a series of high-quality highly-oriented thin films on a Si substrate. A high inverse piezoelectric coefficient (55.7 pm/V) is demonstrated for the thin films. Two types of domain bands can be observed (in the size of a few microns): type-I band tilts ~60° with respect to the horizontal axis, while the type-II band is perpendicular to the horizontal axis. Most of the domain walls (DWs) are 180° DWs for the two bands, while some 109° DWs can also be observed. Interesting, the DWs in type-I band are curved, charged domain walls; while the 180° DWs in type-II band are straight, noncharged domain walls. After applying +20 V for 5 s through a PFM tip, the 180° DWs in type-I band shrinks first, then disconnecting from the band boundary, forming a needle-like domain with a size of ~100 nm. The needle-like domain will extend toward the band boundary after an inverse bias is applied (-20 V), and expand along the band boundary after touching the boundary. Whereas for the type-II domain band, the 180° DWs are more mobile than the 109° domain walls, which displaces ~500 nm after applying +20 V. While such displacement is much shorter after the application of a negative bias for the same duration, starting from the positively poled sample. We hope to spur further interest in the on-chip design of the molecular ferroelectrics based electronic devices.

**Introduction**

Molecular ferroelectrics are a class of ferroelectric materials composed of small organic molecules with weak intermolecular forces such as hydrogen bonding and van der Waals forces [1-4]. They have been considered a promising candidate for applications in flexible electronics, with low manufacturing temperature, low cost, non-toxic elements, high flexibility, and light weight as compared to the traditional polar oxides. Previously, it has been demonstrated that with careful design, molecular ferroelectric materials can achieve high electric properties that are comparable or even higher than oxide ferroelectrics.

Among all the molecular ferroelectric materials, Imidazole perchlorate ($ImClO_4$) have been widely investigated, with high Curie temperature, facile sample preparation, high spontaneous polarization, low coercive field, etc. [5-8] The discovery of ferroelectricity in $ImClO_4$ can be dated back to 2006 [9], where Pajak *et al.* demonstrate a high Curie temperature of 373 K for $ImClO_4$. In 2014, Zhang *et al.* synthesized the $ImClO_4$ thin films which shows superior electromechanical coupling [5]. Li *et al.* revealed the strong electrocaloric strength in $ImClO_4$ thin films that are 13 time higher than the ferroelectric polymers [7]. Moreover, it is also demonstrated that the ferroelectric properties and electric resistivity can be controlled by the proton transfer [8].

These studies enlightened the potential applications of the ImClO4 based ferroelectric materials and devices. Meanwhile, there are still two obstacles towards industrial applications for this system. Firstly, the integration of the molecular thin films in Si-based devices (e.g., transistors, memories, and sensors) is difficult. Unlike the inorganic oxides that can be deposited layer-by-layer on a substrate, the growth of high quality molecular ferroelectric thin films on Si are challenging since the chemical and physical properties of $ImClO_4$ and Si are vastly different. Second, the deterministic control of the ferroelectric domain switching, as well as the fundamental understanding of the domain switching kinetics are lacking, which requires integrated experimental and theoretical efforts.

Herein, we have successfully synthesized high-quality, large-scale, and highly-oriented ImClO4 single crystals using a facile solution-evaporation method. Excellent ferroelectric and piezoelectric properties are obtained for the single crystals. Highly (-110)-oriented $ImClO_4$ thin films are then grown directly on a Si substrate with the sol-gel spin-coating and annealing method, using sodium alginate (SA) as the nucleation additive. A high inverse piezoelectric coefficient (55.7 pm/V) is demonstrated for the thin films. PFM characterizations indicate the formation of two types of domain bands (in the size of a few microns): type-I band tilts ~60° with respect to the horizontal axis, while the type-II band is perpendicular to the horizontal axis. Most of the domain walls (DWs) are 180° DWs for the two bands, while some 109° DWs can also be observed. Interesting, the DWs in type-I band are curved, charged domain walls; while the 180° DWs in type-II band are straight, noncharged domain walls. The switching kinetics for the two bands are further investigated. After applying +20 V for 5 s through a PFM tip, the 180° DWs in type-I band shrinks first, then disconnecting from the band boundary, forming a needle-like domain with a size of ~100 nm. The needle-like domain will extend toward the band boundary after an inverse bias is applied (-20 V), and expand along the band boundary after touching the boundary. Whereas for the type-II domain band, the 180° DWs are more mobile than the 109° domain walls, which displaces ~500 nm after applying +20 V. While the domain wall displacement is much shorter after the application of a negative bias for the same duration, starting from the positively poled sample. The facile synthesis route for thin films on Si and in-depth understanding of the switching kinetics could pave the way towards the on-chip design of the molecular ferroelectrics based electronic devices.

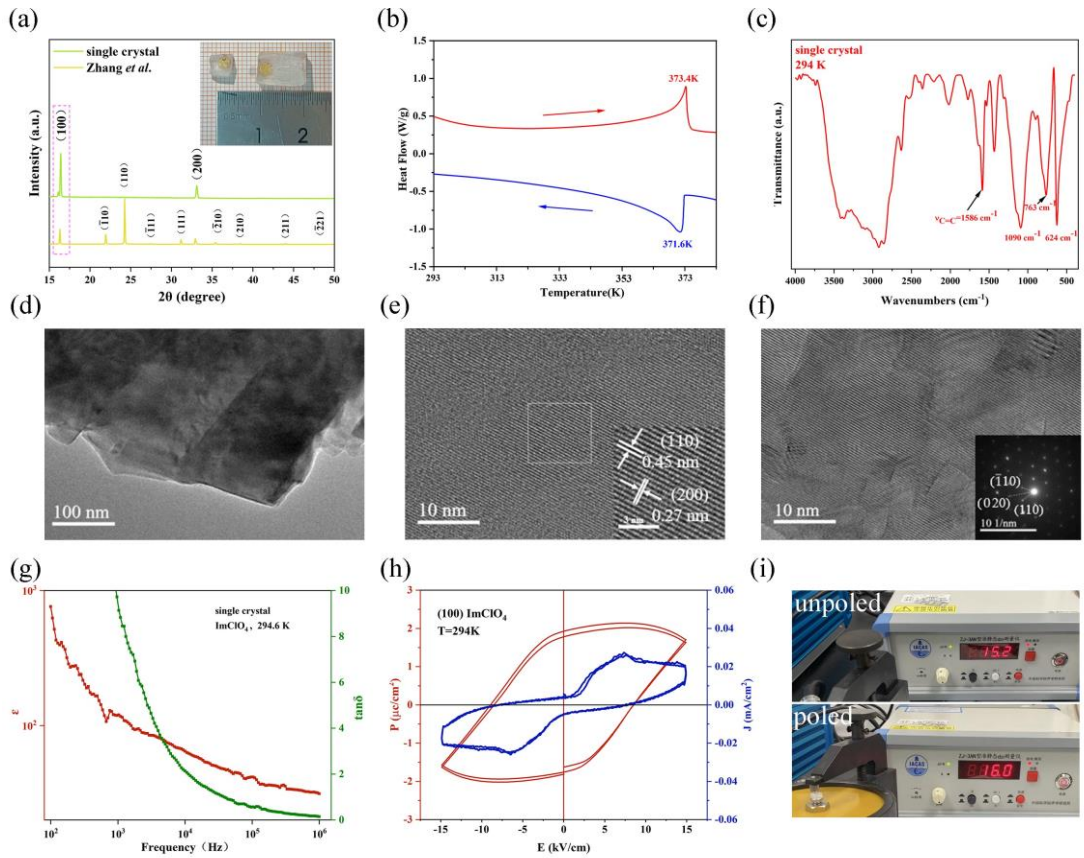

**Fig. 1| Structure characterizations and electric properties of the ImClO$_4$ bulk single crystal.** (a) XRD pattern of ImClO$_4$ single crystal with the crystal morphology. (b) Differential scanning calorimetry (DSC) measurements of ImClO$_4$ single crystal. (c) Fourier Transform Infrared (FTIR) spectra of ImClO$_4$ single crystal. (d)-(f) The high-resolution TEM (HRTEM) images of ImClO$_4$ with a fast Fourier transform of the HRTEM image. (g) The dielectric constant (ε) and the dielectric loss (tanδ) of ImClO$_4$ were measured on the single crystals at various frequencies. (h) The typical polarization-electric field (P-E) hysteresis loops measured on the single crystal with its current density curve. (i) Piezoelectric properties of single crystal samples tested before and after polarization.

The structure characteristics and electrical properties of the as-grown ImClO$_4$ single crystal are given in **Fig. 1,** details of the single crystal growth procedure are introduced in methods. Transparent single crystals with a size up to 1.2×0.8×0.4 cm$^3$ have been successfully synthesized (Fig. 1a). Only two diffraction peaks at 16.40° and 33.12° can be identified, corresponding to the (100) and (200) lattice planes, respectively. This indicates that our sample is highly oriented along [100] direction with good crystallinity, in contrast to the single crystal grown by Zhang *et al.* [5] that is mainly oriented along [110] direction with multiple peaks. The Differential Scanning Calorimetry (DSC) characterization is further performed to unveil the ferroelectric transition temperature, as shown in Fig. 1 (b). Endothermic/exothermic peaks can be seen at 373. 4 K/371.6 K, showing that the ferroelectric phase transition is a first-order type transition with a Curie temperature ~372 K, consistent with the previous report [9]. The FTIR transmission peaks at 624 $cm^{-1}$, 763 $cm^{-1}$, and 1090 $cm^{-1}$ correspond to the out-of-plane vibration, in-plane deformation, and symmetric stretching of the $[ClO_4]^-$ groups, respectively [7]. Whereas the peak at 1586 $cm^{-1}$ represents the symmetric stretching of imidazolium cations from the C=C bond. All in all, we have successfully prepared highly [100]-oriented $ImClO_4$ single crystals, with a facile low-temperature synthesis method.

The high-resolution transmission electron microscope was employed to investigate the morphology and crystal plane of a $ImClO_4$ crystal, as shown in Fig. 1 (d)-(f). Fig. 1 (d) shows the morphology of a $ImClO_4$ single crystal, the black and white contrast with a size of ~100 nm can be seen, corresponding to the different ferroelectric domains separated by a domain wall. A magnified view of the single crystal is presented in Fig. 1 (e), showing the periodic lattice planes, e.g., the (-110)//(200) crystal planes with a lattice spacing of 4.5 Å and 2.7 Å, respectively. This is further confirmed by the selected area electron diffraction (SAED) pattern, where the (200), (-110), and (110) diffraction points can be identified, confirming the orientations of the single crystal (Fig. 1 f).

The electric properties of the single crystal are further measured. Fig. 1 (g) shows the dielectric constant and dielectric loss of $ImClO_4$ single crystal as a function of different frequencies at ambient temperature. As the frequency increases, the dielectric constant and dielectric loss both show a monotonic decreasing trend. From 1 kHz to 1 MHz, the relative dielectric permittivity decreases from 119.86 to 31.3, while the dielectric loss also decreases from 9.13 to 0.14. A rectangular P-E loop with a saturation polarization of 1.81 μC/cm² at 294 K can be obtained with a low coercive field of ~8.5 kV/cm, as shown in Fig. 1 (h). Moreover, the current peaks obtained from current density measurements further verify the existence of ferroelectricity in ImClO₄ single crystals. Furthermore, the unpoled ImClO₄ single crystals possess a high positive piezoelectricity coefficient ($d_{33}$) of 15.2 pC/N at room temperature, which increases slightly to 16 pC/N after poling (Fig. 1i). These measurements demonstrate that our highly (100)-oriented single crystals show superior electric properties.

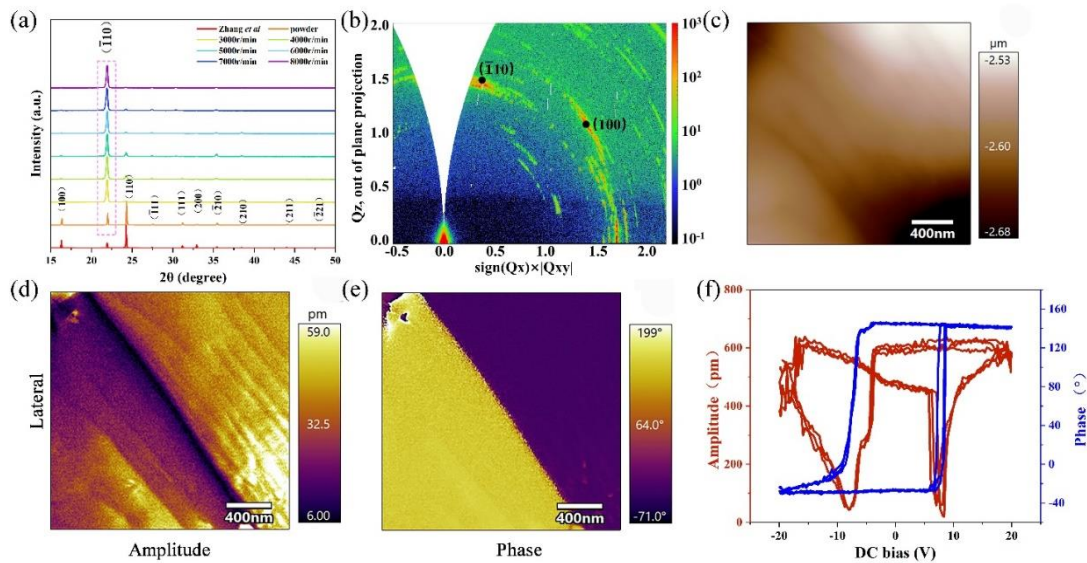

**Fig. 2| Structure characterizations, ferroelectric domain characterizations, and second harmonic generation (SHG) measurements of the highly oriented ImClO₄ thin film.** (a) XRD pattern of ImClO₄ thin film doped with 1 wt% sodium alginate prepared at different rotational speeds. (b) The grazing-incidence wide-angle X-ray scattering (GIWAXS) pattern was obtained from a highly oriented ImClO₄ thin film. (c)-(e) Topography, lateral amplitude, and phase images of PFM for ImClO₄ films. (f) The local PFM hysteresis loops: the amplitude (red) and phase (blue) signals as functions of bias voltage.

In order to better explore the compatibility of molecular ferroelectric materials and silicon-based devices, we prepared ImClO₄ thin films (down to 1 μm) on single-crystal silicon wafers using the sol-gel spin-coating annealing method (details in **methods**). SA is added in the solvent to assist the nucleation process and oriented growth. The structural characterizations are performed for the as-grown thin films. The XRD patterns are shown in Fig. 2 (a), for the ImClO$_4$ films prepared by spin coating at different rotational speeds (3000-8000 r/min), the (-110) diffraction peaks dominate, confirming the highly-oriented growth of the ImClO$_4$ films assisted with SA. The synchrotron-based Grazing Incidence Wide Angle X-Ray Scattering (GIWAXS) are used to evaluate the quality of the thin films (Fig. 2b). It can be seen that the (-110) diffractions peak shows the highest intensity, with a slightly weaker (100) peak. It can be concluded that these films are highly oriented along [-110] direction, while slightly surface reconstruction can be observed with (100) crystal planes.

The domain structure is characterized by piezoelectric force microscope (PFM). Fig. 2 (c) shows the topography of the ImClO$_4$ film prepared by spin coating at 6000 r/min. It can be observed that a region with a flat surface is selected. The lateral amplitude and phase are presented in Fig. 2d-2e, a clear domain wall with 180° phase difference can be identified, with decreasing amplitude in the vicinity of the domain wall. Fig. 2 (f) illustrates the ferroelectric switching with butterfly-shaped amplitude (red) and phase (blue) hysteresis loops. Besides, by calculating through the highest and lowest ends of the butterfly wings, the inverse piezoelectric coefficient of the ImClO$_4$ film is ~55.7 pm/V, which is among the highest reported piezoelectric response values in this system.

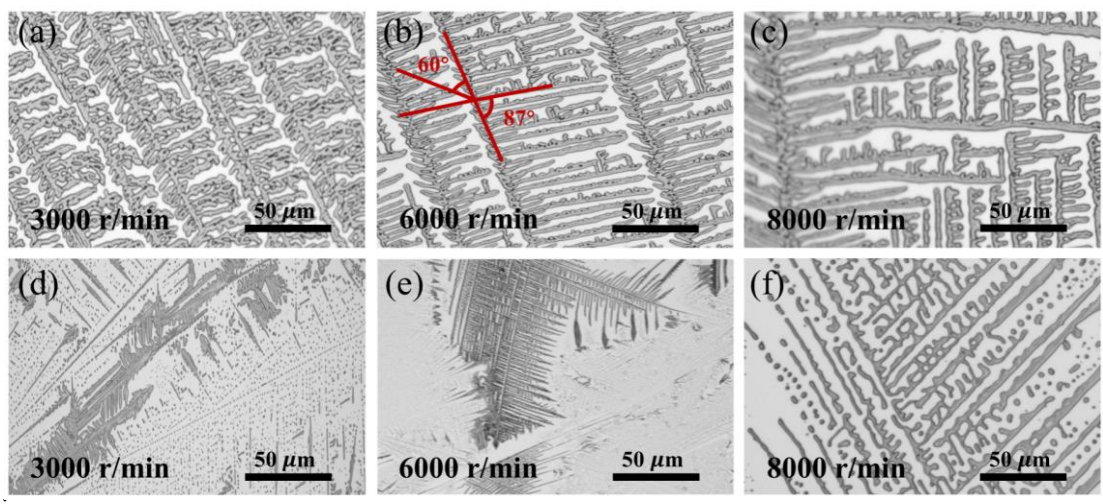

**Fig. 3| Optical microscopy characterization results of the ImClO₄ thin films.** (a)-(c) ImClO₄ morphology of doped 1 wt% sodium alginate at different rotational speeds (3000 r/min、6000 r/min、8000 r/min) under optical microscope. (d)-(f) ImClO₄ morphology at different rotational speeds (3000 r/min、6000 r/min、8000 r/min) under optical microscope without doped 1 wt% sodium alginate.

The surface morphology is characterized by optical microscope, for the samples prepared with and without sodium alginate (SA) additives (**Fig. 3**). As shown in Fig. 3 (a)-(c), herringbone-like morphology can be seen for the ImClO₄ films prepared with SA. Interestingly, with a higher rotation speed, the size of the "bone" increases (~50 μm @3000 r/min, ~100 μm @6000 r/min, and >200 μm @8000 r/min). At a high rotation speed (e.g., 8000 r/min), the side branches can be seen from the branches on the stem. It's worth noting that the angles (**Fig. 3b**) between a stem and the branches are 60° on the left side and 87° on the right side, which is consistent with previous reports. Meanwhile, without the SA additive, various morphologies can be observed, with bubble-like structure, herringbone-like structure, and needle-like structure. These morphologies are irregular in space, and not obvious trend can be seen with different rotation speed. This study indicates that the SA can act as a nucleation additive which provide preferred nucleation sites to homogenize the system during thin film growth, leading to the growth of highly oriented thin films.

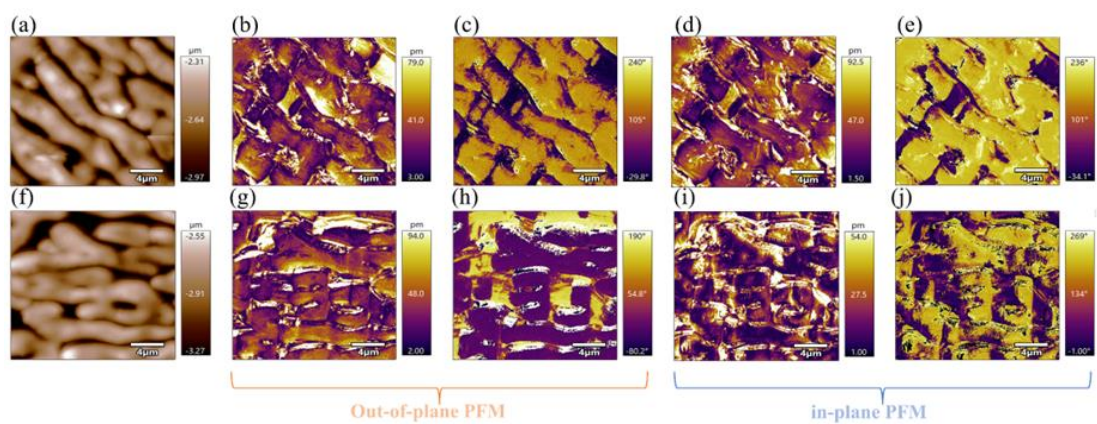

**Fig. 4| The PFM image for ImClO₄ thin films prepared by spin coating at 6000 r/min.** (a) (f) Topography images. (b) (g) Vertical PFM amplitude images. (c) (h) Vertical PFM phase images. (d) (i)

Lateral PFM amplitude images. (e) (j) Lateral PFM phase images.

The polar domains for the SA-assisted ImClO$_4$ thin films are characterized by piezoelectric force microscopy (PFM, **Fig. 4**). Two types of structures can be seen from the topology mapping, corresponding to two types of morphologies (referred as type-I and type-II domains). The topology, out-of-plane amplitude, out-of-plane phase, inplane amplitude, and inplane phase for the type-I domains are shown in Fig. 4 (a)-(e), respectively. In this case, the domain boundaries tilt ~60° with respect to the horizontal axis, while the domain walls are perpendicular to the boundaries. Most domain walls show 180° changes in both inplane and out-of-plane phases, indicating that they are 180° domain walls; while some other domain walls show only 180° shift in out-of-plane phase, indicating that they can be 71° domain walls. In this case, the 180° domain walls dominate. Whereas the type-II domains are oriented parallel or perpendicular to the horizontal axis. Both 180° domain walls and 71° domain walls can be identified from the inplane and out-of-plane amplitude and phase changes. Notably, unlike the first case, the correlation between the inplane and out-of-plane amplitude is weak, showing the formation of pure inplane 180° domain walls, consistent with the domain walls calculated from the phase-field simulations. Furthermore, the population for the 71° domain walls is higher for the second case as compared to the first case.

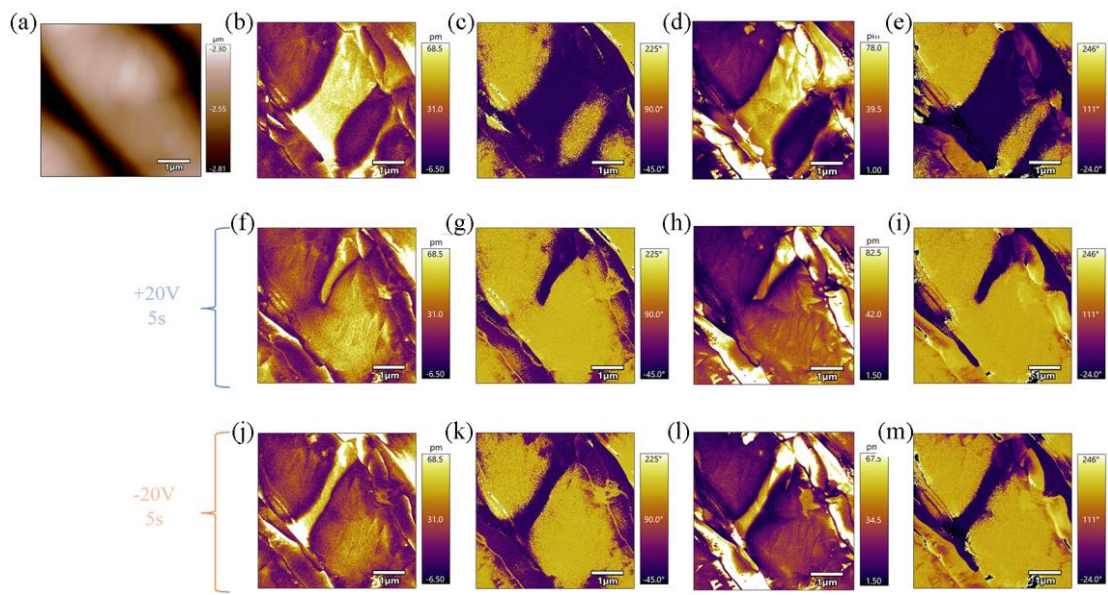

**Fig. 5| The insitu PFM switching for the type-I domains.** (a) Topography images with 2.5 × 2.5 μm². (b)-(e) Initial VPFM amplitude, VPFM phase, LPFM amplitude, LPFM phase images, respectively. (f)-(i) VPFM amplitude, VPFM phase, LPFM amplitude, and LPFM phase images after applying a bias of +20 V for 5 s in the center of the area, respectively. (j)-(m) VPFM amplitude, VPFM phase, LPFM amplitude, and LPFM phase after applying a bias of -20 V for 5 s in the center of the area, respectively.

The insitu switching study for type-I domain walls is performed by applying a local bias through a static PFM tip, as illustrated in **Fig. 5**. Fig. 5 (a) shows the morphology of a selected area (2.5 × 2.5 μm²) of typical type-I domains from Fig. 4 (a), where the domain boundary tilts 60° with respect to the horizontal axis. The initial domain structure can be seen from the PFM images in Fig. 5(b)-5(e), where the 180° domain walls are perpendicular to the boundary, forming band-like superdomain topology. Across the domain wall, the inplane and out-of-plane polarization magnitude is similar. The domain walls are highly curved, indicating that they can be charged domain walls. The formation of charged domain walls in weak ferroelectrics has been investigated previously [10, 11]. After applying a positive bias of 20 V for 5 s in the center of the region, the domain band shrinks gradually, disconnecting from the bottom domain, and forming submicron needle-like domains (Fig. 5f-5i). Whereas the small initial 180° domains have been completely switched. An opposite bias (-20 V) is applied further, after 5 s, the needle domain extends towards the band boundary, while the size of the needle is almost constant. This indicates that the needle edge is a highly charged state that tends to respond quicker to the applied electric field. The whole switching process can be repeated, where the application of positive bias leads to the formation of needle-like domains, while a reverse bias could invert the switching. Notably, when the pulse is kept for a longer time (-20 V, 10 s), the needle will expand along the band boundary where the size can increase to 1 μm.

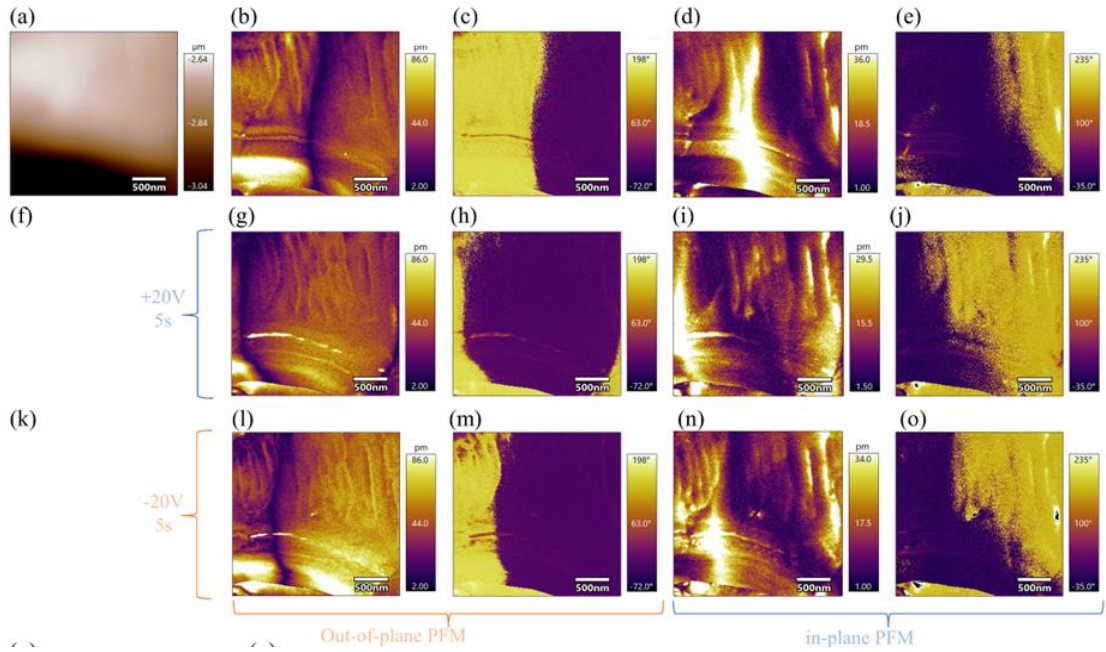

**Fig. 6| The insitu PFM switching for the type-II domains.** (a) Topography images with 2.5 × 2.5 μm². (b)-(e) Initial VPFM amplitude, VPFM phase, LPFM amplitude, LPFM phase images, respectively. (f)-(i) VPFM amplitude, VPFM phase, LPFM amplitude, and LPFM phase images after applying a bias of +20 V for 5 s in the center of the area, respectively. (j)-(m) VPFM amplitude, VPFM phase, LPFM amplitude, and LPFM phase after applying a bias of -20 V for 5 s in the center of the area, respectively.

Next, we proceed to understand the switching kinetics of the type-II domains (**Fig. 6**). The initial PFM images suggest the coexistence of 180° and 109° domain walls that are perpendicular or orients 45° to the horizontal axis (Fig. 6b-6e). The 180° domain wall is very straight, suggesting that it is a non-charged domain wall. After applying +20 V for 5 s, the 180° domain wall moves to the left side, while the 109° domain wall is less mobile, which only extend with the motion of the 180° domain wall (Fig. 6g-6j). Consequently, with the application of a reverse bias for 5 s, the 180° domain wall moves to the right side for 200 nm, much shorter than the initial displacement of 500 nm after applying +20 V (Fig. 6l-6o). This suggests that the field driving motion for the 180° domain wall is not fully reversible, which could cause fatigue for this system.

**Conclusion**

In summary, we have developed a facile solution-evaporation method to grow large scale (~1 cm in size) high quality (100)-oriented ImClO$_4$ single crystals. They exhibit good ferroelectricity (spontaneous polarization ~1.81 μC/cm$^2$ at 294 K), low coercive field (~8.5 kV/cm), high Curie temperature (~372 K), and high piezoelectric coefficient (15.2 pC/N for unpoled crystal).

Then, highly (-110)-oriented ImClO$_4$ thin films are grown on a Si substrate with the sol-gel spin-coating and annealing method, using sodium alginate as the nucleation additive. XRD, optical microscopy, and PFM characterizations are performed to unveil the crystal structure, morphology, piezoelectric, and ferroelectric properties of the thin films. A high inverse piezoelectric coefficient (55.7 pm/V) is obtained in this film. Two types of domain bands (in the size of a few microns) are observed: type-I band tilts ~60° with respect to the horizontal axis, while the type II band is perpendicular to the horizontal axis. Both domain bands are formed mainly by 180° domain walls, with minor 109° domain walls. Most of the domain walls in type-I band are curved, charged domain walls; while the 180° domain walls in type-II band are straight, noncharged domain walls. The switching kinetics for the two bands are investigated with PFM. After applying +20 V for 5 s, the 180° domain walls in type-I band shrinks first, then disconnecting from the band boundary, forming a needle-like domain with a size of ~100 nm. While the smaller domains (~100 nm) are completely switched. The needle-like domain will extend toward the band boundary after an inverse bias is applied (-20 V), starting from the needle tip point, and expand along the band boundary after touching the boundary. Whereas for the type-II domain band, the 180° domain walls are more mobile than the 109° domain walls, which displaces ~500 nm after applying +20 V. While the domain wall displacement is much shorter after application of a negative bias for the same time, starting from the positively poled sample. This study provides a facile synthesis route for the growth of highly-oriented, high quality, and large-scale molecular ferroelectric single crystals and thin films, as well as comprehensive, in-depth understanding of the domain and domain wall kinetics. The thin films are grown directly on a Si substrate, paving the way towards the on-chip design of the electronic devices.

**Acknowledgement**

ZH would like to acknowledge a financial support from the National Natural Science Foundation of China (No. 12174328). A startup grant from Zhejiang University is also acknowledged.

**Methods**

**Crystal growth and film preparation**

(100)-oriented molecular ferroelectric single crystals with sizes up to 1.2×0.8×0.4 cm were prepared by slow evaporation of equimolar imidazole (500 mmol) and perchloric acid (500 mmol) solutions at room temperature. To get $ImClO_4$ films, $ImClO_4$ crystals obtained by evaporating the solution and sodium alginate were dissolved in deionized water and stirred for 24 hours to form a solution, wherein the proportion of sodium alginate mixed was 1 wt%. After using a plasma cleaning machine to carry out hydrophilic treatment on the surface of the single-crystal silicon substrate, the previously prepared solution was spin-coated on the substrate surface evenly at a speed of 3000-8000 r/min through a homogenizer. A series of molecular ferroelectric $ImClO_4$ films of different thicknesses can be prepared by drying the films in an oven at a low temperature of 80°C for 1 h and annealing the films again at 120°C for 1h. The specific synthesis process of crystal growth and film preparation can be found in Figure S1 in the Supporting Information.

**DSC, SHG, XRD and GIWAX measurements**

Differential scanning calorimetry (DSC) measurements were carried out by using a DSC Q100 instrument under the nitrogen atmosphere, where the single crystal of $ImClO_4$ (2.94 mg) was heated and cooled with a rate of 10 K/min in the temperature ranges of 288–394 K. The X-ray diffraction (XRD) data of ImClO$_4$ single crystal were obtained from Rigaku fabricated Miniflex-600c in the 2θ range of 15°–50° with a step size of 0.02°. The grazing incident XRD (GIXRD) data of ImClO$_4$ film at different rotational speeds were collected to detect the intensity of crystal planes on the Bruker D8 Discover using Cu Kα radiation (λ = 1.54059 Å) with a scan rate of 10° min$^{-1}$. Powder X-ray diffraction (PXRD) was obtained from Rigaku fabricated Miniflex-600c and was analyzed by depositing ImClO$_4$ powder in a glass substrate with 0.5 mm depth, and the detecting angles started from 2θ = 15 ° to 2θ = 50 °, with 0.02 ° increment. GIWAX measurements were performed at Xeuss 3.0 with high-brightness micro-focus spot solid-state Cu target light source (λ = 1.542 Å)

**Dielectric and ferroelectric measurements**

The instrument BALAB DMS2000 was used to detect the frequency-dependent dielectric constants of ImClO$_4$ single crystals under the frequency range from 100 Hz to 1 MHz with an applied electric field of 1 V. The longitudinal piezoelectric coefficient ($d_{33}$) was tested by the quasi-static meter (ZJ-3A, Chinese Academy of Sciences, China) under the low-frequency constant force (0.25 N) at room temperature. The P-E loops and current density of ImClO$_4$ single crystals were carried out using a ferroelectric test system (Premier II, Radiant Technologies Inc., USA). The ferroelectric polarization imaging and local switching measurements on the bulk crystal surface were carried out using a PFM (MFP-3D, Asylum Research).

**FTIR and Optical microscopic measurements**

The flourier-transformed infrared spectroscopy (FTIR, Vertex 70) under transmission mode and attenuated total reflection-infrared (ATR-IR) mode were utilized to characterize the bulk crystals and films, respectively. The optical microscopic images were obtained by a probe stage (HCP421V-MPS).

**Declare of Interest**

The authors declare no competing interest.